\documentstyle[12pt]{article}

\newlength{\dinwidth}
\newlength{\dinmargin}
\begin{document}

\def\bold#1{\setbox0=\hbox{$#1$}%
     \kern-.025em\copy0\kern-\wd0
     \kern.05em\copy0\kern-\wd0
     \kern-.025em\raise.0433em\box0 }
\def\slash#1{\setbox0=\hbox{$#1$}#1\hskip-\wd0\dimen0=5pt\advance
       \dimen0 by-\ht0\advance\dimen0 by\dp0\lower0.5\dimen0\hbox
         to\wd0{\hss\sl/\/\hss}}
\def\lq{\left [}
\def\rq{\right ]}
\def\LL{{\cal L}}
\def\VV{{\cal V}}
\def\AA{{\cal A}}
\def\MM{{\cal M}}

\newcommand{\be}{\begin{equation}}
\newcommand{\ee}{\end{equation}}
\newcommand{\bea}{\begin{eqnarray}}
\newcommand{\eea}{\end{eqnarray}}
\newcommand{\nn}{\nonumber}
\newcommand{\dd}{\displaystyle}
\newcommand{\bra}[1]{\left\langle #1 \right|}
\newcommand{\ket}[1]{\left| #1 \right\rangle}
\thispagestyle{empty}
\vspace*{1cm}
\rightline{BARI-TH/94-180}
\rightline{June 1994}
\vspace*{2cm}
\begin{center}
  \begin{Large}
  \begin{bf}
$D^*$ radiative decays and strong coupling of heavy mesons with soft pions in
 a  QCD relativistic potential model\\
  \end{bf}
  \end{Large}
  \vspace{8mm}
  \begin{large}
P. Colangelo $^{a,}$ \footnote{e-mail address:
COLANGELO@BARI.INFN.IT},  F. De Fazio $^{a,b}$, G. Nardulli $^{a,b}$\\
  \end{large}
  \vspace{6mm}
$^{a}$ Istituto Nazionale di Fisica Nucleare, Sezione di Bari, Italy\\
  \vspace{2mm}
$^{b}$ Dipartimento di Fisica, Universit\'a
di Bari, Italy \\

\end{center}
\begin{quotation}
\vspace*{1.5cm}
\begin{center}
  \begin{bf}
  ABSTRACT
  \end{bf}
\end{center}
\vspace*{0.5cm}
In the framework of a QCD inspired relativistic potential model, we evaluate
radiative decay rates of heavy mesons and their coupling with soft pions.
The agreement with the existing experimental data  is satisfactory. In
the limit $m_Q \to \infty$ one obtains results in agreement with the Heavy
Quark Effective Theory  and is able to predict the values of the relevant
couplings; in particular, for the scaled $B^*B \pi$ strong coupling constant
$g$, we find that the non relativistic constituent quark model prediction $g=1$
is modified, by the inclusion of the relativistic effects due to the
light quarks,
to the value $g=1/3$, in agreement with recent QCD sum rules determination.
\noindent
\end{quotation}

\newpage
\baselineskip=18pt
\setcounter{page}{1}

In this letter we wish to analyze two problems. The first one is the study of
the flavour conserving $D^*$ decays:

\be D^* \to D \gamma \hskip 3 pt , \label{eq : 1}\ee

\be D^* \to D \pi \hskip 3 pt . \label{eq : 2}\ee

\noindent Even though we do not posses a full set of experimental data yet,
results from  CLEO II collaboration \cite{CLEO} already yield useful
information on (\ref{eq : 1}) and (\ref{eq : 2}) and provide constraints on
the theoretical models. The second problem addressed
by the present letter is the determination of the strong coupling $g_{D^*D\pi}$
, defined by:

\be <D^0(k) \; \pi^+(q)|D^{*+}(p,\epsilon)>=g_{D^*D \pi}\epsilon^\mu q_\mu
\hskip 3 pt \label{eq : 3} \ee

\noindent in the limit $m_c \to \infty$. This coupling has been considered by a
number of authors \cite{defazio, gatto, casalbuoni, grozin}. In fact, it is of
interest in chiral effective theories of heavy mesons that describe the strong
interactions of heavy $(Q{\bar q})$ mesons with chiral Nambu-Goldstone bosons
or light vector mesons, as well as their couplings to weak and electromagnetic
currents \cite{wise, cho, amundson, georgi, burdman, yan}. \par
In the  $m_c \to \infty$ limit $g_{D^* D \pi}$ can be written as
follows \cite{nussinov}:

\be g_{D^* D \pi}={2 m_D \over f_\pi} g \label {eq : 4} \ee

\noindent and one can show that, in the constituent quark model $g \simeq 1$
\cite{cinesi, isgur} \footnote{The value quoted in ref.\cite{cinesi} ($g=1$)
stems from the knowledge of the spin configurations of heavy mesons in the non
relativistic quark model, while the authors in ref. \cite{isgur} find a
slightly different
value: $g \simeq 0.8$, obtained in a calculation considering mock mesons (see
references therein).}.
However, recent analyses indicate smaller values;
for example from the study of the semileptonic  $D \to \pi \ell \nu_\ell$
decay one estimates $g=0.4$ (this value is obtained using a chiral effective
theory and performing the limit $m_c \to \infty$ \cite{defazio},
\cite {casalbuoni}). Also recent QCD sum rules analyses of the coupling
(\ref{eq : 3}) in the
heavy quark infinite mass limit point to small values $(g $ in the range $ 0.
2-0.4 )$ \cite{gatto}. \par
Prompted by these results one may wonder if this departure from the naive quark
constituent model might be understood as a consequence of the neglect of the
relativistic motion of the light antiquark ${\bar q}$ inside the meson $D$.
We can test this simple explanation of a low value for $g$ in
a well defined model for the heavy mesons interactions, based
on a constituent quark picture of the hadrons; the strong interaction between
the quarks is described by a QCD inspired potential
and the relativistic effects due to the kinematics are included
in the wave equation \cite{cea, pietroni, tedesco}. \par
Let us briefly describe the main features of this model. The heavy hadrons
$D_a$ and $D^*_a$ made up by the quark $Q$ and the antiquark ${\bar q_a}$ are
described by the states:

\bea |D_a(p)>=&i&{\delta_{\alpha \beta} \over \sqrt{3}}
{\delta_{rs} \over \sqrt{2}}
\int d \vec{k} \; \psi (\vec{k}+x\vec{p}, -\vec{k}+(1-x)\vec{p})\times
\nonumber
\\
 &\times& b^{\dag}(\vec{k} + x \vec{p}, r, \alpha) \;
d^{\dag}_a(-\vec{k}+(1-x)\vec{p}, s, \beta)|0> \hskip 3 pt, \label{eq : 5}
\\
|D^*_a(p,\epsilon)>=&&{\delta_{\alpha \beta} \over \sqrt{3}}
{(-\epsilon^\mu \sigma_\mu)_{rs} \over \sqrt{2}} \int d \vec{k} \;
\psi (\vec{k}+x\vec{p}, -\vec{k}+(1-x)\vec{p})\times \nonumber \\
 &\times& b^{\dag}(\vec{k} + x \vec{p}, r, \alpha) \;
d^{\dag}_a(-\vec{k}+(1-x)\vec{p}, s, \beta)|0> \hskip 3 pt, \label{eq : 6}
\eea

\noindent where $\alpha$ and $\beta$ are colour indices, $r$ and $s$ are spin
indices, $b^{\dag}$ and $d^{\dag}_a$ are creation operators of the quark $Q$
and the antiquark ${\bar q_a}$, carrying momenta $\vec{k} + x \vec{p}$ and
$-\vec{k} + (1-x) \vec{p}$ respectively. The meson wave function
$\psi(\vec{k} + x\vec{p} , -\vec{k} + (1-x) \vec{p})$ satisfies the
Salpeter wave function \cite{salpeter}\footnote[1]{This equation arises from
the bound-state Bethe Salpeter equation by considering the instantaneous time
approximation and restricting the Fock space to the $Q{\bar q}$ pairs; for more
details see \cite{pietroni}.}, which includes relativistic effects due to the
kinematics explicitly:
\bea
\Big\{ \sqrt{(\vec{k}+x \vec{p})^2 + m_Q^2} &+&
\sqrt{[-\vec{k}+(1-x) \vec{p}]^2 + m_{q_a}^2}-\sqrt{m_D^2+\vec{p}^2} \Big\}
\psi( \vec{k} + x \vec{p}, -\vec{k} + (1-x) \vec{p}) \nonumber \\
&+& \int d \vec {k^{\prime}} V(\vec{p}, \vec{k}, \vec{k^{\prime}})
  \psi (\vec{k^{\prime}}, \vec{p}-\vec{k^{\prime}})=0 \label{eq : 7} \eea

\noindent which is valid in a moving frame where the meson $D$ (or $D^*$),
having mass $m_D$, has momentum $\vec{p}$; the wave function $\psi$ is
normalized as follows:

\be {1 \over (2\pi)^3} \int d \vec{k} |\psi|^2=2\sqrt{m_D^2+\vec{p}^2} \hskip 3
pt , \label{eq : 8} \ee

\noindent whereas the instantaneous potential $V$ coincides, in the meson rest
frame, with the Richardson potential \cite{richardson} and in the $r$-space
takes the form:

\be V(r) ={ 8 \pi \over 33-2n_f} \Lambda \left( \Lambda r
-{f(\Lambda r) \over \Lambda r} \right) \hskip 3 pt , \label{eq : 9} \ee

\noindent where $\Lambda$ is a parameter, $n_f$ is the number of flavours and
\footnote[2]{This potential grows linearly when $r \to \infty$ and follows QCD
predictions for small $r$.}:
\be  f(t)={4 \over \pi} \int_0^\infty dq
{sin ( qt) \over q} \Bigg[ {1 \over ln(1+q^2)} -{1 \over q^2} \Bigg]
\hskip 3 pt . \nonumber \ee
 In order
to avoid unphysical singularities \cite{nardulli}, we assume that $V(r)$, near
the origin, is constant:

\be V(r)=V(r_M) \hskip 1 cm \left( r \leq r_M={\lambda  \over 3 m_D }{4 \pi
\over 3} \right)
\hskip 3 pt . \label{eq : 10} \ee

\noindent The values of the parameters, as obtained by fits to meson masses,
are as follows:
$m_u=m_d=38 \hskip 3 pt MeV$; $m_s=115 \hskip 3 pt  MeV$, $m_c=1452
\hskip 3 pt MeV$, $m_b=4890 \hskip 3 pt MeV$, $\Lambda=397 \hskip 3 pt MeV$,
$\lambda=0.6$. We note that fits of the heavy meson masses do not constrain
light quark masses uniquely; in particular one could obtain similar results for
the meson masses by considering , e.g. $m_d=m_u \simeq 100 \hskip 5 pt MeV$ and
including a small and negative constant term $V_0$ in the potential.\par
In order to compute the amplitudes for the decays (\ref{eq : 1}) and
(\ref{eq : 2}), i.e. $g_{D^* D \pi}$ in (\ref{eq : 3}) and $g_V$ given by (
$e$=electron charge):

\be <D^+(k)| J^{e.m.}_\mu|D^{*+}(p,\epsilon)>=g_V \; e \;
\epsilon_{\mu \nu \alpha \beta} \; \epsilon^\nu k^\alpha p^\beta
\hskip 3 pt , \label{eq : 11} \ee

\noindent we have to express the currents in terms of quark operators.
We write:

\be J^{e.m.}_\mu=Q_{ij} \delta_{\alpha \beta} \int {d \vec{q} \; d
\vec{q^{\prime}} \over (2 \pi)^3}
\left[{m_i m_j \over E_i(\vec{q}) E_j(\vec{q^{\prime}})} \right]^{1/2}
{\bar q_i} (\vec{q}, r, \alpha) \gamma_\mu
q_j(\vec{q^{\prime}},s,\beta) \hskip 3 pt , \label{eq : 12} \ee

\noindent where $E_i(\vec{q})=\sqrt{m_i^2+\vec{q}^2}$, $Q_{ij}$ is the quark
charge matrix and $q_i$, $q_j$ represent the usual quark field operators. \par
We also consider the axial vector current $A_\mu$, which is obtained from
(\ref{eq : 12}) by substituting $\gamma_\mu \to \gamma_\mu \gamma_5$ and
$Q_{ij}$ with the appropriate flavour matrix. The matrix element of the axial
current can be written as follows:

\bea <D^0(k) | A_\mu | D^{*+}(p, \epsilon) > =-i \{&\epsilon_\mu&(m_{D^*}+m_D)
A_1(q^2)-{\epsilon \cdot q \over m_D + m_{D^*}}(p+k)_\mu A_2(q^2)  \nonumber \\
&-& {\epsilon \cdot q \over q^2} 2 m_{D^*} q_\mu [A_3(q^2)-A_0(q^2)] \} \hskip
 3 pt , \label{eq : 13} \eea

\noindent where $2m_{D^*} A_3=(m_D + m_{D^*})A_1+(m_{D^*}-m_D)A_2$ ($q=p-k$).
\par
Taking the derivative of $A_\mu$, we obtain $(J_5=i{\bar d} \gamma_5 u)$:

\be (m_u+m_d)<D^0(k)|J_5|D^{*+}(p,\epsilon)>=-i(\epsilon \cdot q)\; 2m_{D^*}
A_0(q^2) \hskip 3 pt . \label{eq : 14} \ee

\noindent For small $q^2$ the matrix element on the l.h.s. of (\ref{eq : 14})
is dominated by the $\pi^+$ pole, therefore we obtain, for $q^2$ small:

\be g_{D^* D \pi} ={m_\pi^2 -q^2 \over m_\pi^2 } {2 m_{D^*} \over f_\pi} A_0(
q^2) \hskip 3 pt , \label{eq : 15} \ee

\noindent which gives the $\pi D D^*$ coupling in the chiral limit $(q^2=0)$:

\be g_{D^* D \pi} = {2 m_{D^*} \over f_\pi} A_0(0) \hskip 3 pt .
\label{eq : 16} \ee

We can now compute $g_V$ in (\ref{eq : 11}) and $A_0(0)$ in (\ref{eq : 16}) by
our model; we closely follow the approach described in ref. \cite{tedesco} to
which we refer the interested reader for further details. We obtain:

\be g_V={e_Q \over \Lambda_Q} + {e_q \over \Lambda_q} \hskip 3 pt ,
\label{eq : 21} \ee

\noindent where:

\bea \Lambda_Q^{-1} &=& {1 \over 2m_D} \int_0^\infty dk |\tilde{u}(k)|^2
 {1 \over E_Q} \left( 1-{k^2 \over 3 E_Q(E_Q+m_Q)} \right) \nonumber \\
\Lambda_q^{-1}&=& {1 \over 2m_D} \int_0^\infty dk |\tilde{u}(k)|^2
{1 \over E_q} \left( 1- {k^2 \over 3
E_q (E_q+m_q)} \right) \hskip 3 pt , \label{eq : 17} \eea

\noindent where $e_Q$, $e_q$ are the heavy and light quark electric charges,
 $E_j=\sqrt{k^2 +m_j^2}$ ($j=q,Q$ and we put $m_u=m_d=m_q$) and $\tilde{u}(
k)$ is related to the wave function $\psi$ of eq. (\ref{eq : 7}) with
$\vec{p}=0$ by the equation:

\be \tilde{u}(k)={k \; \psi(k) \over \sqrt{2} \pi } \hskip 3 pt .
\label{eq : 18} \ee

\noindent $\tilde{u}(k)$ can be obtained by solving (\ref{eq : 7}) numerically
by the Multhopp method \cite{multhopp}. In computing eqs. (\ref{eq : 21},
\ref{eq : 17}) we have put $m_D=m_{D^*}=2.025 \hskip 5 pt GeV$ which is the
theoretical value obtained by solving eq. (\ref{eq : 7}). As for the other
channels, we have put: $m_B=m_{B^*}=5.33 \hskip 3 pt GeV$
$m_{D_s}=m_{D^*_s}=2.05 \hskip 3 pt GeV$ $m_{B_s}=m_{B^*_s}=5.366 \hskip 3 pt
GeV$.
$\Lambda_Q$ and $\Lambda_q$ are the mass parameters whose
values are reported in Table I. It may be useful to compare (\ref{eq : 21})
with the result of the constituent quark model \cite{eichten}, where:
$\Lambda_q=m_d=m_u=0.335$ $ GeV$ or  $\Lambda_q=m_s=0.45$ $ GeV$ and
$\Lambda_Q=m_c=1.84$ $ GeV$ for the charm case.
\par
Let us now consider $g_{D^* D \pi}$ as given by eq. (\ref{eq : 16}).
Let us observe that in the present approach one obtains the value of the form
factors at $q^2=q^2_m=(m_{D^*}-m_D)^2$
 (or $(m_{B^*}-m_B)^2$), whereas we actually need $A_0(0)$; since for small
$q^2$ $A_0
(q^2)$ has a pole at the pion mass, this approximation could in principle
introduce relevant differences, expecially for the $D$ case. However we observe
that by choosing the potential (\ref{eq : 9}), we have neglected spin-spin
terms, which corresponds to the approximation $m_D=m_{D^*}$; for consistency we
have therefore to put $q^2_m=0$ as well, which is what we shall do.
Using the
tecniques of ref. \cite{tedesco} we obtain the following equation for the form
factor $A_1$ in (\ref{eq : 13}):

\be  A_1(q^2_m)={1 \over 4m_D} \int_0^\infty dk |\tilde{u}(k)|^2
     \left( {E_q + m_q \over E_q} \right) \left[1 - {k^2 \over 3(E_q +m_q)^2}
\right]  \hskip 3 pt , \label{eq : 22} \ee

\noindent Moreover, one has:

\be A_3(q^2)={m_D+ m_{D^*} \over 2m_{D^*}} A_1(q^2) -
             {m_D- m_{D^*} \over 2m_{D^*}} A_2(q^2) \hskip 3 pt ;
\label {eq : a_3} \ee

\be A_3(0)=A_0(0) \hskip 3 pt ; \label{eq : a_0} \ee

\noindent therefore, in the limit $m_D=m_{D^*}$ one gets:

\be A_3(0)=A_1(0) \hskip 3 pt , \label{eq : a_1} \ee

\noindent provided that $A_2(0)/m_Q^2$ goes to zero when $m_c \to \infty$,
which we have tested numerically for $m_Q \gg m_c$ using relations in ref.
\cite{tedesco}  \footnote{We have used the analogous of eq.(36) of
\cite{tedesco}; one obtains eq.
(\ref{eq : a_1}) provided that the difference between $x_{D^*}$ and $x_D$ (see
(\ref{eq : 5}), (\ref{eq : 6}) for their definition), does not go zero
more rapidly than $1/m_Q$.}.\par
A numerical analyses gives the result:

\be A_0(0) = 0.40 \hskip 2 cm (D  \; \; case)  \label{eq : 24} \ee
\be A_0(0) = 0.393 \hskip 2 cm (B  \; \; case)  \label{eq : 25} \ee

 Our results
are therefore, in the chiral limit :

\be g_{D^*D\pi}\simeq 12.3 \hskip 3 pt , \label{eq : 26} \ee
\be g_{B^*B\pi}\simeq 31.7 \hskip 3 pt , \label{eq : 27} \ee

\noindent which shows a deviation of only $2 \hskip 3 pt \%$ from the scaling
result $g_{D^*D\pi}/g_{B^*B\pi}=m_D/m_B$.\par
Using the results of eqs. (\ref{eq : 21}), (\ref{eq : 26}), (\ref{eq : 27}) and
Table I, we are able to compute the widths for the possible strong and
radiative $D^*$ and $B^*$ decay channels. These results are reported in Table
II together with the available experimental information. We observe that these
results compare favourably with the experimental data
whenever they are available; they are also
in agreement with an analysis of these decays performed by us in the framework
of HQET for mesons containing one heavy quark \cite{defazio}. It should be
observed that in the our calculation we had no free parameters.
\par

Let us now discuss the determination of the scaled constant $g$ defined by
(\ref{eq : 4}):

\be g=A_0(0)=A_1(0)=
{1 \over 4 m_D} \int_0^\infty dk |\tilde{u}(k)|^2 {E_q + m_q
\over E_q} \left[ 1-{k^2 \over 3 (E_q+m_q)^2} \right] \hskip 3 pt .
\label{eq : 28} \ee

It is interesting to consider the non-relativistic limit, where: $E_q \simeq
m_q \gg k$. In this limit one obtains:

\be g={1  \over 2 m_D} \int_0^\infty dk |\tilde{u}(k)|^2=1  \label{eq : 29} \ee

\noindent because of the normalization condition (\ref{eq : 8}). Eq. (\ref{eq :
29}) reproduces the well known constituent quark model result
\cite{cinesi, isgur}.\par Let us now take in (\ref{eq : 28}) the limit
$m_q \to 0$, which is possible since we work in the chiral limit and there is
no restriction to the values of $m_q$ in the Salpeter equation (incidentally
our fit of the meson masses is obtained with the rather small value $m_q=38$
$MeV$). In this case, we obtain:

\be g={1 \over 3} \hskip 3 pt . \label{eq : 30} \ee

\noindent It is worth to stress that the strong reduction of the value of $g$
from the naive non relativistic quark constituent model value $g=1$
 (eq.(\ref{eq : 29}))to the
result (\ref{eq : 30}) has a simple explanation in the effect of the
relativistic kinematics taken into account by the Salpeter equation.
Furthermore, our asymptotic value of $g$ is in the range of values
($g=0.2-0.4$) determined in \cite{gatto} by means
of QCD sum rules in the limit of infinite heavy quark mass and assuming
massless light quarks, that is in the same approximation that we have used
in (\ref{eq : 30}). \par

In conclusion, the potential model  has allowed us to
describe the radiative decays of heavy mesons, obtaining a rather good
agreement with  other similar approaches and with existing experimental data.
Moreover, it has given us the possibility to give a simple explanation,
based on the relativistic kinematics, for the small value of the strong
coupling constant $g$ of (\ref{eq : 3}), (\ref{eq : 4}) predicted by QCD sum
rules in the $m_Q \to \infty$ and its deviation from the non relativistic quark
model result.
\vskip 1cm
\noindent {\bf Acknowledgments\\}
\noindent We thank N.Paver for interesting discussions.

\newpage
\begin{center}
  \begin{Large}
  \begin{bf}
  Table Captions
  \end{bf}
  \end{Large}
\end{center}
  \vspace{5mm}
\begin{description}
\item [Table I]
Mass parameters of eq. (\ref{eq : 21}) (values in $GeV$).
\end{description}
  \vspace{10mm}
\begin{description}
\item [Table II] Radiative and hadronic decay widths of $B^*$ and $D^*$ mesons.
\end{description}

\newpage

\begin{table}
\begin{center}
\begin{tabular}{l c c}
 & {\bf Table I} &  \\ & & \\
 \hline \hline
Decay mode  & $\Lambda_Q$ & $\Lambda_q$ \\ \hline
$ D^* \to D \gamma $ & 1.57 & $0.48 $ \\
\hline
$ D^*_s \to D_s \gamma $ & 1.58 & $0.497 $ \\
\hline
$ B^* \to B \gamma   $ & 4.93 & $ 0.59$ \\
\hline
$ B^*_s \to B_s \gamma $ & 4.98 & $0.66 $ \\
\hline \hline

\end{tabular}
\end{center}
\end{table}

\newpage
\begin{table}
\begin{center}
\begin{tabular}{l c c }
 & {\bf Table II} &  \\ & & \\
 \hline \hline
Decay rate/ BR & theory & experiment \\ \hline
$\Gamma(D^{*+})$ & $46.21  \hskip 3 pt KeV$ & $<$ 131 \hskip 5 pt KeV
\cite{accmor} \\ \hline
$ BR(D^{*+} \to D^+ \pi^0)$ & $31.3 \%$ & $30.8 \pm 0.4 \pm 0.8 \%$ \\
\hline
$BR(D^{*+} \to D^0 \pi^+)$ & $67.7 \%$ & $68.1 \pm 1.0 \pm 1.3 \%$ \\
\hline
$BR(D^{*+} \to D^+ \gamma)$ & $1.0 \%$ & $1.1 \pm 1.4 \pm 1.6 \%$ \\
\hline \\ \hline
$\Gamma(D^{*0})$ & $41.6 \hskip 3 pt KeV$ &  \\ \hline
$ BR(D^{*0} \to D^0 \pi^0)$ & $50.0 \%$ & $63.6 \pm 2.3 \pm 3.3 \%$ \\
\hline
$BR(D^{*0} \to D^0 \gamma)$ & $50.0  \%$ & $36.4 \pm 2.3 \pm 3.3 \%$ \\
\hline  \\ \hline
$\Gamma(D^*_s)=
\Gamma(D^*_s \to D_s \gamma)$ & $ 0.382  \hskip 3  pt KeV$ &  \\ \hline
\\ \hline
$\Gamma(B^{*-})=\Gamma(B^{*-} \to B^- \gamma)$ & $0.243
\hskip 3  pt KeV$ &  \\ \hline \\  \hline
$\Gamma(B^{*0})=\Gamma(B^{*0} \to B^0 \gamma)$ & $9.2 \hskip 3 pt 10^{-2}
\hskip 3  pt KeV$ &  \\ \hline \\ \hline
$\Gamma(B^*_s)=\Gamma(B^*_s \to B_s \gamma)$ & $8.0 \hskip 3pt 10^{-2}
\hskip 3  pt KeV$ &  \\ \hline  \hline

\end{tabular}
\end{center}
\end{table}
\vspace{20 mm}
\vspace{20 mm}

\end{document}